\documentclass{aa}
\usepackage{graphicx}
\begin{document}

\title{Discovery of a peculiar DQ white dwarf
\thanks{Based on observations made with the William Herschel Telescope
operated on the island of La Palma by the Isaac Newton Group in
the Spanish Observatorio del Roque de los Muchachos of the
Instituto de Astrofisica de Canarias.}\fnmsep
 \thanks{Based on observations made with the Italian Telescopio Nazionale Galileo (TNG) operated on the
 island of La Palma by the Centro Galileo Galilei of the INAF (Istituto Nazionale di Astrofisica) at
 the Spanish Observatorio del Roque de los Muchachos of the Instituto de Astrofisica de Canarias.}
}

%\subtitle{}
\author{D. Carollo,
    \inst{1}
    S.T. Hodgkin,
    \inst{2}
      A. Spagna,
      \inst{1}
      R.L. Smart,
      \inst{1}
      M.G. Lattanzi,
      \inst{1}
      B.J. McLean,
      \inst{3}
    D.J. Pinfield
      \inst{4}
        }

%          \fnmsep\thanks{Just to show the usage
%          of the elements in the author field}

%          }
   \offprints{Daniela Carollo \\ \tt e-mail: carollo@to.astro.it}
   \institute{INAF, Osservatorio Astronomico di
   Torino, I-10025 Pino Torinese, Italy\\
    \and Cambridge Astronomical Survey Unit, Institute of
    Astronomy, Madingley Road, Cambridge, CB3 0HA, UK\\
    \and Space Telescope Science
    Institute (STScI), Baltimore, MD 21218, USA\\
    \and Astrophysics Research Institute, Liverpool John Moores University,
         Birkenhead, CH41 1LD, UK\\
    }

   \date{Received 23 January 2002/Accepted 2 August 2002}

\abstract{We report the discovery of a new carbon rich white dwarf
that was identified during a proper motion survey for cool white
dwarfs based on photographic material used for the construction of
the Guide Star Catalog II. Its large proper motion ($\mu\simeq
0.48$ arcsec/yr) and faint apparent magnitude ($V\simeq 18.7$)
suggest a nearby object of low luminosity. A low-resolution
spectrum taken with the William Herschel Telescope clearly shows
strong C$_2$ Deslandres-d'Azambuja and Swan bands, which identify
the star as a DQ white dwarf. The strength of the
Deslandres-d'Azambuja bands and the depression of the continuum in
the Swan-band region are signs of enhanced carbon abundance for
the given $T_{\rm eff}$.  Comparison of our spectrophotometric
data to published synthetic spectra suggests 6000 K $ < T_{\rm
eff} <$ 8000 K, although further analysis with specialized
synthetic models appear necessary to derive both  $T_{\rm eff}$
and chemical composition. Finally, the range of spatial velocity
estimated for this object makes it a likely member of the halo or
thick disk population. }

\titlerunning{Discovery of a peculiar DQ star }
\authorrunning{Carollo et al.}

\maketitle

%
%________________________________________________________________
\section{Introduction}

Star GSC2U J131147.2+292348 was identified during a proper motion
survey for cool halo white dwarfs (WDs) based on photographic
material used for the construction of the Second Guide Star
Catalogue (GSC-II) (see, e.g., Lasker et al.\ 1995, McLean et al.\
2000). The object is located near the North Galactic Pole (NGP) at
$l\simeq 61^\circ$, $b \simeq 85^\circ$, is fast moving
($\mu\simeq 0.48$ arcsec~yr$^{-1}$), and faint ($V\simeq 18.7$),
as expected for a low luminosity object in the solar neighborhood.
An accurate check on the SIMBAD database revealed that the star is
not in the NLTT catalogue (Luyten 1979) but, quite surprisingly,
is listed as a quasar candidate (object OMHR 58793) by Moreau \&
Reboul (1995), who measured an UV excess but did not detect any
proper motion.

%__________________________________________________________________
\section{Observations and Data Analysis}

\begin{table*}
  \centering
  \caption{GSC2 plate material used for the astrometry and the photographic photometry of the new DQ white dwarf.}

  \begin{footnotesize}
\begin{tabular}{rcccccl} \hline\hline
  % after \\ : \hline or \cline{col1-col2} \cline{col3-col4} ...

  Field &  Survey & Center (J2000) & Epoch  & Pixel & Color & Emulsion + Filter \\
  \hline

  XJ443 & POSS-II & 13:04:14.7 +29:48:37 &  1995.234 & 15 $\mu$m & $B_J$ & IIIaJ + GG385 \\
  XP443 & POSS-II & 13:04:15.2 +29:48:42 &  1993.288 & 15 $\mu$m & $R_F$ & IIIaF + RG610 \\
  XI443 & POSS-II & 13:04:20.7 +29:44:17 & 1991.299 & 15 $\mu$m & $I_N$ & IV-N + RG9 \\
  N322  & Quick V & 13:06:56.6 +29:13:25 & 1983.294 & 25 $\mu$m & $V_{12}$ & IIaD+Wratten 12 \\
  XE322 & POSS-I  & 13:06:55.5 +29:13:25 &  1955.288 & 25 $\mu$m & $E$ & 103a-E + red plexiglass \\
  XO322 & POSS-I  & 13:06:56.1 +29:13:24  & 1955.288 & 25 $\mu$m & $O$ & 103a-O unfiltered \\
 \hline

\end{tabular}
 \end{footnotesize}
  \label{plates}
\end{table*}

\subsection{Astrometry and photometry}
 Our material consists of Schmidt plates from the Northern photographic surveys (POSS-I, Quick V and POSS-II) carried out at the Palomar Observatory (see Table 1). All plates were digitized at STScI utilizing modified PDS-type scanning machines with 25 $\mu$m square pixels (1.7 $''$/pixel) for the first epoch plates, and 15 $\mu$m pixels (1 $''$/pixel) for the second epoch plates (Laidler et al. 1996). These digital copies of the plates were initially analyzed by means of the standard software pipeline used for the construction of the GSC-II. The pipeline performs object detection and computes parameters and features for each identified object.  Further, the software provides classification, position, and magnitude for each object by means of astrometric and photometric calibrations which utilized the Tycho2 (H{\o}g et al.\ 2000) and the GSPC-2
(Bucciarelli et al.\ 2001) as reference catalogs.  Accuracies
better than 0.1-0.2~arcsec in position and 0.15-0.2 mag in
magnitude are generally attained.

\begin{figure}
%   \centering
%  \includegraphics[width=\textwidth]{poss1and2.eps}
  \includegraphics[width=9cm]{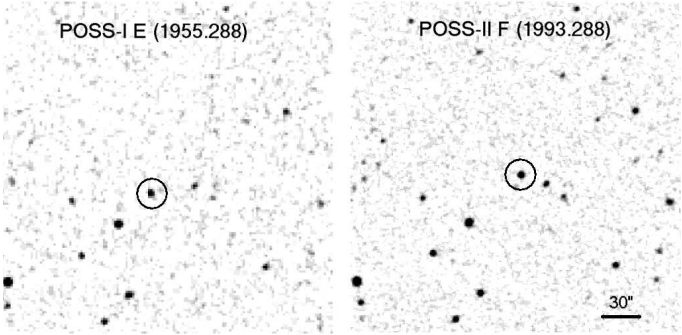}
% \rule{0.3pt}{2cm}% line thickness, height of picture
  \caption{First epoch (POSS-I, XE322) and second epoch (POSS-II, XP443) plates in the direction of the newly discovered WD, the encircled star near the field center. The large relative motion of the object is evident.}
              \label{finding_chart}
\end{figure}

\begin{table}[b]
  \centering
\caption[]{Astrometry and photometry of GSC2U J131147.2+292348.
The position was determined from plate XJ443 (epoch 1995.234,
equinox J2000), while all of the available plates were used for
the proper motions. The error of the photographic photometry is
better than 0.2 mag (1$\sigma$).}

\begin{tabular}{ccccc} \hline
\hline $\alpha$ (h m s)& $\delta$ (d m s)&
$\mu_{\alpha}\cos\delta$ & $\mu_{\delta}$\\
 (J2000) & (J2000)  &   (arcsec/yr)  &  (arcsec/yr) \\
\hline
 13 11 47.21 & +29 23 48.0 & $-0.382 \pm 0.002$ & $0.286 \pm 0.005$ \\
\hline
\end{tabular}
%\label{Astro}
%\end{table}

\vspace{0.20cm}

%\begin{table}
%  \centering
%\caption[]{Photometry derived from photometric system of POSS-II
%plates}

\begin{tabular}{cccc} \hline
\hline
  %% after \\ : \hline or \cline{col1-col2} \cline{col3-col4} ...
%\multicolumn{4}{c}{CCD} & \multicolumn{2}{c}{POSS-II} \\
    $ B_J $ & $ V_{12} $ & $ R_F $ & $ I_N $ \\
 \hline
 19.6 & 18.7 & 18.1 & 17.5  \\
 \hline
\end{tabular}

\vspace{0.20cm}

\begin{tabular}{ccc}
 \hline \hline
 $ J $ & $ H $ & $ K_{s} $  \\
 \hline
 $17.48\pm 0.05$ & $17.13\pm 0.10$ & $17.08\pm 0.12$   \\
 \hline

\end{tabular}
%\label{Phot}
\label{parameters}
\end{table}

Star GSC2U J131147.2+292348 was part of the sample of WD
candidates discovered after screening the high proper motion stars
found in survey field 443 (Table 1). These were selected on the
basis of their relative proper motions as derived by applying the
procedure described in Spagna et al.\ (1996) to just the POSS-II
plates, spanning $\sim 4$ years.  The finding charts in Figure 1
show the high proper motion of this object.

The astrometry and photometry of GSC2U J131147.2+292348 are given
in Table 2. The position refers to the epoch of the most recent
plate (XJ443), while the accurate proper motion was computed by
combining the image locations of the star as measured on the 6
different plates of Table 1, which span $\sim 40$ years. The
photographic magnitudes are given in the natural photometric
system of the POSS-II and Quick-V plates as defined by the
emulsion-filter combinations in Table 1.

In particular, the transformation between the photographic and
Johnson $V$ is $V_{12}=V - 0.15 (B-V)$ according to Russell et
al.\ (1990). Also, recently acquired NIR images \footnote{Taken
with the NICS camera on the 3.6-m TNG telescope on La Palma}
provided the J, H, K$_{\rm s}$ magnitudes in Table 2. Finally,
Moreau \& Reboul (1995) published the values $U\simeq 19.15$ and
$V\simeq 19.10$. Note that their visual magnitude is fairly
consistent with our $V_{12}$, considering the above color
transformation and the errors of the photographic photometry.

%%%%%%% ************************************************** %%%%%
%%%%%%% ++++++++++++++++++++++++++++++++++++++++++++++++++ %%%%%

\subsection{Spectroscopy}
Spectroscopy of GSC2U J131147.2+292348 was obtained on the night
of 2001 January 29 using the intermediate dispersion
spectrographic and imaging system (ISIS) on the 4.2-m William
Herschel Telescope on the island of La Palma. The 5700 \AA {}
dichroic was used to split the light and feed to the blue and red
arms of the spectrograph.

We used the R158B grating on the blue arm, which gave a nominal
dispersion of 1.62 \AA/px and useful wavelength coverage from 3200
to 5700 \AA. (The dichroic cuts in at wavelengths $> 5700$ \AA,
and at short wavelengths, the sensitivity falls off with the
quantum efficiency of the detector). On the red arm, we used the
R158R grating to give a nominal dispersion of 2.9 \AA/px covering
from 5500 to 8000 \AA. A blocking filter (GG495) was also used on
the red arm to cut out second order blue light. A 30-minute
exposure was made using a 1-arcseconds slit. Subsequent exposures
were taken of the spectrophotometric standards Feige 67 and Feige
34 to enable flux calibration of the primary target. We took arc
lamp exposures to enable wavelength calibration and tungsten lamp
exposures for the pixel-to-pixel sensitivity variation and enable
flat fielding.

The data were reduced within the IRAF environment, following
standard procedures.  No attempt was made to correct for
extinction, both standards and targets were measured with an
airmass $\le 1.1$. Observations were made with a slit width of
1.02 arcseconds, which corresponds to 4 detector pixels in the
blue, i.e. a dispersion of 6.5 \AA {} per resolution element. For
the red arm, the pixel scale is 0.36 arcseconds per pixel, leading
to a resolution element of size 3 pixels, i.e. a resolution of 8.2
\AA. The blue and red arm spectra have been gaussian smoothed at
these resolutions.

Good agreement between the red and blue arm spectra was found in
the overlap region, with fluxes agreeing to better than 10\% in
the range 5600-5700 \AA.

\begin{figure*}
%\resizebox{16cm}{!}{\includegraphics[angle=-90,width=6cm]{spectrum2.eps}}
\resizebox{16cm}{!}{\includegraphics[angle=-90,width=6cm]{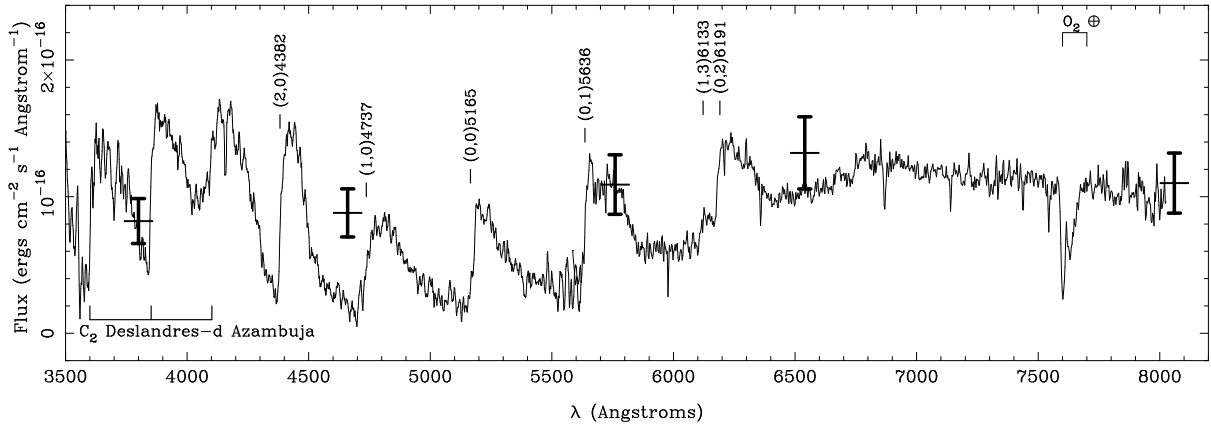}}
 \centering
 %%%%%\rule{0.3pt}{2cm}% line thickness, height of picture
   %%\includegraphics{empty.eps}
   %%\includegraphics{empty.eps}
  \caption{The WHT optical spectrum of GSC2U J131147.2+292348.
  Vertical marks indicate the locations of the strong C$_{2}$
  Deslandres-d'Azambuja and Swan bands, and of the telluric O$_{2}$.
  The crosses refer to the fluxes (with $\pm 20 \%$ error bars)
  derived from the B$_{J}$, V$_{12}$,
  R$_{F}$, I$_{N}$ photographic photometry of Table 2 and the U mag
  from Moreau and Reboul (1995).}
              \label{spectra}%
\end{figure*}

%*********************************************************************%%
%%%%%*****************************************************************%%

\section{On the nature of GSC2U J131147.2+292348}

The flux-calibrated spectrum of GSC2U J131147.2+292348 is shown in
Figure 2. The signal-to-noise is around 10 for the whole spectrum,
increasing slightly to the red. This noise level is clearly
visible in the spectrum, and limits our ability to detect weak
features.

The crosses in Figure 2 represent the fluxes at different
effective wavelengths as derived from the $B_{J}$ , $V_{12}$,
$R_{F}$, and $I_{N}$ photographic magnitudes in Tab. 2. The
ultraviolet flux was derived from the photographic U magnitude of
Moreau et al.\ (1995). The agreement appears reasonably consistent
with the 10\% and 20\% accuracy levels of the flux-calibrated
spectroscopy and the photographic photometry, respectively.

The spectrum appears dominated by strong absorption bands due to
C$_{2}$ molecules. The four Swan bands with bandheads at
$\lambda=$ 4382, 4737, 5165, and 5636 \AA~ are clearly identified,
along with the less common Swan band at 6191 {\AA}. In addition,
strong Deslandres and d'Azambuja (D-d'A) absorption bands are also
present in the blue part of the spectrum at 3600, 3852, and 4102
\AA. These bands have been observed in the spectra of WDs with
carbon rich atmospheres (DQ WDs) and temperatures\footnote{This
temperature seems to be the lower limit for DQ stars, and might be
associated with the transition of C$_2$ into C$_2$H molecules
(Bergeron et al.\ 2001).} above 6500 K. Finally, the spectrum in
Figure 2 shows an evident depression of the continuum in the Swan
band region between 4500 and 6200 {\AA}.

%%%%%%%%%%%%%%%%%%%%%%%%%%
%%%%%%%%%%%%%%%%%%%%%%%%%%

The spectral energy distribution (SED) of DQ stars changes with
$T_{\rm eff}$ and carbon abundance as shown by the model
atmosphere spectra presented in Koester et al.\ (1982) and Wegner
\& Yackovich (1984). Figure~5 of Wegner \& Yackovich gives and
indication on what to expect for different combinations of $T_{\rm
eff}$ and C:He abundance. Swan bands are generally present, while
D-d'A bands start to become visible in models with C:He$\ga
10^{-6}$ at $T_{\rm eff}\simeq 6600$~K and with C:He $\ga 10^{-2}$
at $T_{\rm eff} \simeq 10\,000$ K.

A SED with C$_2$ bands similar in strength to those observed in
our spectrum requires a much enhanced C:He ratio for the given
$T_{\rm eff}$. This can be seen by comparing the models in
Figure~5 of Wegner \& Yackovich with those in their Figures~2 and
~3. At temperatures between 6000~K and 7000~K, deep absorption
bands are produced with C:He~$\approx 10^{-4}$. At $T_{\rm eff} =
8000$~K, carbon abundance has to increase to a rather extreme
value, C:He=0.9, for the simultaneous presence of strong D-d'A
{\it and} Swan bands in the synthetic SED (bottom panel of
Figure~3 of Wegner \& Yackovich). This model bears the most
resemblance with the spectrum of our WD, however, it does not show
any evidence of the continuum depression seen in the observed
spectrum. Theoretical evidence that such depression of the
continuum emission could occur is provided in Koester et al.\
(1982). Their Figure~1 displays theoretical C$_2$ spectra at
$T_{\rm eff} = 8000$~K and increasingly higher C:He ratios. The
effect is to boost band strengths, thus depressing the continuum
in the Swan-band region.

% Moreover the model with $T_{\rm eff} = 8000$~K just examined
%predicts a flux ratio of the emission at blue and red wavelengths
%which seems not consistent with the observed SED.

%consistent with the appearance of the C$_2$ band systems observed
%in the spectrum of our WD,
%%%%%%%%%%%%%%%%%%%%%%%%%%%%%%%%%%%%%%%%%%%%%%%%%%%
%%%%%%%%%%%%%%%%%%%%%%%%%%%%%%%%%%%%%%%%%%%%%%%%%%%

Although the models with $T_{\rm eff} = 8000$~K just examined seem
consistent with the appearance of the C$_2$ band systems observed
in the spectrum of our WD, the relative flux at blue wavelengths
(below $\sim$ 4100 \AA) is probably too high compared to the
observed SED in Figure 2. In this regard, an attempt to find a
black body compatible with the observed spectrum at $\lambda >
7000$~\AA, the NIR fluxes from our JHK$_s$ magnitudes, and with
the blue peaks in the D-d'A region, resulted in a black-body
temperature of $\sim 6000~$K. (Note that in this case the
depressed continuum occurs in the region of maximum black-body
emission.)

From the discussion above, it is evident that much is still to be
learned about the properties of this new DQ star, and the reliable
determination of its temperature and chemical composition must
await more detailed atmosphere models. Also, improved spectral
coverage in the UV, below 3500~\AA, would probably be of help in
better constraining model calculations.

%%%%%%%%%%%%%%%%%%%%%%%%%%%%%%%%%

Finally, an approximate photometric parallax for GSCU
J131147.2+292348 was estimated from the absolute magnitudes of
theoretical models of non-DA stars. From the values in Tables 2
and 4 of Bergeron et al.\ (1995) for pure helium atmosphere WDs
and averaging the distance moduli computed for the IJHK$_s$ bands
(which are not affected by the strong C$_2$ absorption bands) we
estimate the distances $d\approx 70$, 80, and 90 parsecs for
$T_{\rm eff}=6000$~K, 7000~K, and 8000~K, respectively.

This distance interval corresponds to a range of tangential
velocity $V_{\rm tan}=4.74 \cdot \mu \, d \simeq
160$-200~km~s$^{-1}$ and galactic components\footnote{Assuming a
solar motion of $U_\odot=10.00$ km~s$^{-1}$ and $V_\odot=5.25$
km~s$^{-1}$, as from Dehnen \& Binney (1998). The (U,V) components
are computed from $\mu_\alpha \cos\delta$, $\mu_\delta$ only.
However, given the high galactic latitude of this star ($b\simeq
85^{\circ}$),  the unknown $V_r$ component would contribute less
than 4\% and 7.6\% to the U and V values, respectively.} with
respect to the LSR from $(U,V)\simeq (-148.1, +9.6)$  to
$(U,V)\simeq (-193.3, +10.8) $ km~s$^{-1}$, for $d=70$~pc and
90~pc, respectively. These relatively high values are not
consistent (3$\sigma$) with the velocity distribution of the thin
disk, while they are consistent with the kinematics of the halo or
thick disk stellar population\footnote{Here, we have adopted the
velocity ellipsoids $(\sigma_U,\, \sigma_V,\, \sigma_W;\,
v_a)=(34, 21, 18; 6)$ km~s$^{-1}$ and $(61, 58, 39; 36)$
km~s$^{-1}$ for the thin and thick disk respectively (Table 10.4
of Binney \& Merrifield 1998). The halo ellipsoid $(\sigma_U,\,
\sigma_V,\, \sigma_W;\, v_a)=(160, 89, 94; 217)$ km~s$^{-1}$ is
from Casertano, Ratnatunga \& Bahcall (1990). Note that these
kinematics parameters are still not well established. In
particular the estimated (U,V) components would result consistent
with the halo kinematics, but only marginally with the thick disk
parameters, recently derived by Chiba and Beers (2000).}.

%Note that the velocity ellipsoids are still not well defined. In
%particular, the estimated (U,V) components would result consistent
%with the kinematics of the halo, but only marginally with that of
%the thick disk, recently derived by Chiba and Beers (2000).}

%Using the more recent evaluations of $(\sigma_U,\, \sigma_V,\,
%\sigma_W;\, v_a)$ from Chiba \& Beers (2000), the estimated (U,V)
%components are consistent only with the halo kinematics.}.

%Using the more recent evaluations of $(\sigma_U,\, \sigma_V,\,
%\sigma_W;\, v_a)$ from Chiba \& Beers (2000), the estimated (U,V)
%components are only marginally consistent with the thick disk
%kinematics .}.

%Note that the thick disk parameters are still controversial. In
%particular, the estimated (U,V) components  would result very
%marginally consistent with the thick disk kinematics from Chiba \&
%Beers (2000) who claim $(\sigma_U,\, \sigma_V,\, \sigma_W;\,
%v_a)=(46\pm 4, 50\pm 4, 35\pm 3; 20\pm 5)$ km~s$^{-1}$.}.

\section{Conclusions}
We have discovered a new carbon rich white dwarf (DQ), which shows
very strong C$_2$ Deslandres-d'Azambuja and Swan bands. To the
best of our knowledge, no other object is known today which such a
strong simultaneous evidence of the two molecular band systems
associated with C$_2$.

Comparisons to published synthetic spectra suggest  6000 $ <
T_{\rm eff} < $ 8000~K, while a black-body fit to the observed
fluxes at $\lambda >$ 7000~\AA, and to the peaks below $\sim$ 4100
{\AA} supports the possibility that $T_{\rm BB}\sim$ 6000~K.
Therefore, it is evident that the reliable determination of
temperature and chemical composition of GSCU J131147.2+292348 must
await more detailed atmosphere model calculations. Anyhow, it is
likely that the carbon abundance in the atmosphere of this WD is
significantly enhanced compared to other known DQ stars of similar
temperature.

A photometric distance of 70-90 parsecs has been estimated, which
implies a relatively large spatial velocity and makes this new DQ
white dwarf a likely member of the halo or thick disk population.
Of course, a direct determination of the distance will be the only
way to derive model independent absolute magnitude and kinematics
for this object.

\begin{acknowledgements}

We are indebted to the referee, U. Heber, for his valuable
comments and suggestions that were essential for the proper
interpretation of our observations. The constant support of our
GSC2 collaborators B.\ Bucciarelli, J.\ Garcia, V.\ Laidler, C.\
Loomis, and R.\ Morbidelli is acknowledged. And thanks go also to
A.\ Boden and R.\ Cutri who reprocessed their 2MASS frames to look
for this object. The GSC~II is a joint project of the Space
Telescope Science Institute and the Osservatorio Astronomico di
Torino. Space Telescope Science Institute is operated by AURA for
NASA under contract NAS5-26555. Current participation of the
Osservatorio Astronomico di Torino is supported by the Italian
National Institute for Astrophysics (INAF). Partial financial
support to this research comes from the Italian CNAA and the
Italian Ministry of Research (MIUR) through the COFIN-2001
program. STH and DJP acknowledge the financial support of the
Particle Physics and Astronomy Reasearch Council of the United
Kingdom. This research has made use of the SIMBAD database,
operated at CDS, Strasbourg (France).

\end{acknowledgements}

\end{document}